# RESEARCH

# Addressing the unmet need for visualizing Conditional Random Fields in Biological Data

William C. Ray[1,2*]
, Samuel L. Wolock[1]
, Nicholas W Callahan[2]
, Min Dong[3]
, Q. Quinn Li[3]
, Chun Liang[3]
, Thomas J Magliery[2]
 and Christopher W. Bartlett[1,2]

## Abstract

**Background:** The biological world is replete with phenomena that appear to be ideally modeled and analyzed by one archetypal statistical framework - the Graphical Probabilistic Model (GPM). The structure of GPMs is a uniquely good match for biological problems that range from aligning sequences to modeling the genome-to-phenome relationship. The fundamental questions that GPMs address involve making decisions based on a complex web of interacting factors. Unfortunately, while GPMs ideally fit many questions in biology, they are not an easy solution to apply. Building a GPM is not a simple task for an end user. Moreover, applying GPMs is also impeded by the insidious fact that the "complex web of interacting factors" inherent to a problem might be easy to define and also intractable to compute upon.

**Discussion:** We propose that the visualization sciences can contribute to many domains of the bio-sciences, by developing tools to address archetypal representation and user interaction issues in GPMs, and in particular a variety of GPM called a Conditional Random Field(CRF). CRFs bring additional power, and additional complexity, because the CRF dependency network can be conditioned on the query data.

**Conclusions:** In this manuscript we examine the shared features of several biological problems that are amenable to modeling with CRFs, highlight the challenges that existing visualization and visual analytics paradigms induce for these data, and document an experimental solution called StickWRLD which, while leaving room for improvement, has been successfully applied in several biological research projects.
  Software and tutorials are available at http://www.stickwrld.org/

## Background

Many biological domains are foundationally based in the study of complex systems of interacting parts. Unfortunately, working biological researchers are caught in a "Chicken and Egg" situation, where modeling approaches that can appropriately represent the complexity, aren't available for lack of tools that support their creation, and there are no tools to support complex model creation because, due to scarcity and difficulty in creation, there is little demand for the models.

Our goal in this manuscript is to catalog the necessary and sufficient features of a visualization or visual analytics system that enables development of useful statistical models of these interactions, and to demonstrate that such a system provides significantly improved insight into biological domains where current methods fail. Herein we document the variety of complex interactions that are critical components of usefully powerful models in many biological systems, outline the characteristics of statistical models that are appropriate for these systems, itemize the requirements for a visualization system intended to support development of such statistical models, and demonstrate that a prototype visual analytics system that addresses these requirements, provides novel and powerful insights into significant and challenging biological problem domains.

Complex Networks of Interacting Features Abound in Biology

Proteins are molecular machines composed of a limited number of basic building blocks, assembled in a myriad of combinations and orders. Not only is the order of assembly important, but for appropriate function, the way that each of the building blocks fits together and interacts with its many spatially proximal (and not necessarily sequentially proximal) neighbors is critical. To make accurate predictions about how a change – a mutation – to a protein will affect its function, requires examining how that change will fit, and function, with many other building blocks in that specific protein. Genomic studies face similar challenges. The panoply of differences between one genome and another, ultimately make each individual distinct, but few of the the differences – inherited Single Nucleotide Polymorphisms, or de-novo mutations – act alone. Instead it is the combinations and mutual interactions of these differences that, in concert, determine the final phenotypic expression of each individual's genomic blueprint. On a larger scale, the complex interplay of

*Correspondence: ray.29@osu.edu
[1]Nationwide Children's Hospital, 575 Children's Crossroad, 43215, Columbus, OH, USA
[2]The Ohio State University, 100 W. 18th Ave, 43210, Columbus, OH, USA
Full list of author information is available at the end of the article normally commensal flora and fauna that inhabit the body is responsible for maintaining a dynamic polymicrobial homeostasis in the gut, mouth, nose, and elsewhere across the body, and minor perturbations to the supportive, competitive, antagonistic or symbiotic relationships amongst the microbial populations are the cause of many infectious diseases.

[Figure 1 about here.]

In each of these cases, domain researchers wish to understand how the system works, by cataloging the observable features of many individuals. From these observations, statistical models are built, that can for example, predict the likelihood that a newly observed individual is a member of the same population that defined the model. Alternatively they can be used to predict the likelihood that, if modified at some feature, a member of the population will still remain a member. In most domains the current state of the art is to build these models as though the features are statistically independent – despite a widespread understanding that this is not appropriate. This happens because there simply isn't a good, accessible way for the domain researchers to define appropriate statistical models that account for the dependencies. It is harder to find biological domains where this situation is not true, than additional domains where it is the standard.

An example of the type of data under consideration, and several canonical summarizations of this data are shown in Figure 1. These data are prototypical of any collection of ordered categorical data: each row $i$ contains a vector of categorical values representing one individual in the training set; each column $j$ contains the categorical value assigned to each individual, for some specific feature; each letter $C_{i,j}$ is simply a single-character symbol denoting the categorical value possessed by individual $i$, for some feature $j$. In practice, sequences in real biological domain problems can be a few hundred positions in length, and might require representing a few dozen different categories. While considerably larger domain problems do exist, in our experience we have found that being able to work with 500 positions and 26 categories has been sufficient to address the large majority of questions in several diverse domains.

Useful and Appropriate Statistical Models must Incorporate Interactions

In all of these domains, the basic data are often represented as sequences, but are fundamentally about networks at the functional level. As a result, the most appropriate statistical models that can aid in understanding the data, and in making predictions about it, will be network-based, rather than sequence-based models. Recent interest in building statistical models based on weighted networks of interacting features holds great promise for these domains. With some variation amongst different graphical probabilistic model designs, the prototypical GPM encodes the marginal distribution of categories observed for each feature using weighted nodes in a graph, and the joint distribution of co-occurring features using weighted edges. Given these weights, which are annealed towards optimal values in a "parameter estimation" step based on training data, the GPM can then produce scores for new observations by integrating across the nodes and edges that those observations select.

Formally, this is to say that generalized GPMs calculate, based on a model-specific encoding of training data, $P(Y_1...Y_n|X_1...X_m)$, for a set of labels $Y$, and a set of observations $X$, where some or all of the elements of $Y$, may also be elements of $X$. Disguised by this description is the detail that GPMs do not treat $X_{1..m}$ as independent. Instead they also encode all pairwise, or potentially higher-order tuple, combinations of elements of $X$. In the specific varieties of GPMs in which we are interested, the pairwise combinations can have weights that are themselves conditional on the actual observations at each element of $X$.

In less formal terms, GPMs can calculate the probability that some collection of features $Y_{1..n}$ are a good fit for the training data, taking into account not just the individual fit of each $Y_i$ to the training data, but also (because $Y$ may overlap $X$) the fit of each $Y_i$ in the specific context of the other observed features in $Y$.

This algorithmic process is an excellent match for what the real world is doing, when it integrates across, for example, the positional and interaction characteristics of a changed amino acid in a protein, to determine the relative activity of a mutant protein compared to the wild-type original. Biology does not evaluate the acceptability of the changed amino acid simply based on the characteristics of the protein family, but rather it evaluates it in the context of both the family characteristics, and of all of the other amino acids in that specific protein and how it has addressed the family needs. Unfortunately, despite the surprising parallels between the algorithmic form, and physical reality, these models have seen limited practical use in the bio/life sciences.

This failure can primarily be laid at the feet of two issues that have restricted the use of graphical probabilistic models largely to theory rather than practical application. The first is that GPMs require, a-priori, a network of features on which to compute statistics, and defining this network for anything beyond trivial data, is beyond the means of most domain researchers. This is especially true if the network connections and weights that must be computed upon, are dependent on the content of the data being analyzed. The second is that even when a realistic network of interactions can be intuited by domain scientists, there is no guarantee that a GPM based upon such a network can be tractably built. In biological domains where the basic understanding is that "everything is connected with everything else at least some level", it is far too easy to build networks with intractably conflicting dependency loops in the network definition.

### Visualization Tools for Building such Models must Represent Interactions with Adequate Detail

Both of these problems can be addressed, if not eliminated, by visualization and visual analytics. However, no current tools provide an appropriate view of the complexity of the data that is necessary for this work. Standard approaches to network visualization are inadequate for several reasons. Chief amongst these, is the conditional existence and weight of network edges, dependent on the data. However, other issues also exist. The prototypical "node" in these domains is some measurable biological feature, such as the nucleotide in a particular position in a gene. The "edges" reflect interactions between that nucleotide and its neighbors. Because the interactions depend on the identity of the nucleotide found in that position in a specific instance of the gene, the edges, and edge weights between a node and other nodes, are dependent on the value found at the node.

*It is important to understand that the bio/life-sciences need is not simply cataloging the strongest of these edges, but rather understanding the patterns and larger networks of the edges, including conditional features of those networks. In practice it is frequently loosely clustered groups of weak, conditional dependencies, that are more important for the domain scientists to understand, than the stronger singular dependencies within the data.*

We could encode this as a vast number of alternative graphs, and select amongst them based on the data, however, less traditional graph formalisms enable this data to be encoded more intuitively. Because there is a fixed set of possible nucleotides that might occur at any node, one can model each node as containing a fixed set of weighted sub-nodes, with each of the dependent edges connecting appropriate sub-nodes from one node to another. As a result, any node $j$ can be connected to another node $k$ by multiple different weighted edges (possibly by the entire weighted bipartite graph between the subnodes of $j$ and the subnodes of $k$).

[Figure 2 about here.]
[Figure 3 about here.]

Graphically, we can represent this structure as shown in Figures 2 and 3.

Formally, this suggests that our data is most appropriately modeled using either multigraphs, or metagraphs[1]. There are features that appear typical in the biological problems, however, that restrict the models to special cases of these formalisms; most specifically the restricted (typically identical) set of sub-nodes available in each node, the omission of edges between nodes and subnodes, and the omission of edges (because the subnodes are mutually exclusive categories within the nodes) between subnodes within the same node. As a result, general tools for multigraphs and metagraphs are unlikely to be optimal for addressing these problems.

**User Requirements**

From the data shown in Figure 1, the working researcher wants (and needs) to understand:

1. The sequential order and relative location in the sequence, of each feature.
2. The marginal distribution of each nucleotide (category) in each sequential position – i.e. the sequences predominantly contain a **C** or **G** symbol in the first position (Figure 2:A, node 1, yellow and green circles), with few **A**s or **T**s, the second position contains an almost equal distribution of **A**s, **G**s and **C**s, with slightly fewer **T**s (Figure 2:A, node 2, similar sized red, green and yellow circles, slightly smaller blue circle).
3. The joint distribution of each possible pair of nucleotides as observed in the training sequences – i.e. a **G** at position 6, almost universally co-occurs with a **C** at position 9 (Figure 3, blue arrow between node 6, subnode **G** and node 9, subnode **C**); **C** at position 6 universally occurs with a **G** at position 9, a **G** at 7, co-occurs with a **C** at 8, etc.
4. When the joint distribution is predictable from the marginal distributions (implying independence), and when the joint distribution differs from the expected distribution (implying dependence).
5. The localized and distributed patterns of the marginal distributions, and interdependent joint distributions, across the entire sequence space – i.e. there are simultaneous dependencies between $G_6$ and $C_9$, $C_6$ and $G_9$, $A_6$ and $T_9$, $T_6$ and $A_9$, $G_7$ and $C_8$, $C_7$ and $G_8$, $A_7$ and $T_8$, and $T_7$ and $A_8$ (Figure 3, blue, magenta, red, brown and grey arrows between subnodes of nodes 6, 7, 8 and 9) which implies a biological feature called a "stem loop" structure. A biological expert end-user would choose to retain these dependencies in the model, regardless of their edge weights. There is also a quite interesting set of dependencies between different triples of nucleotides in columns 2, 3, and 4, knowledge of which is critical to understanding the biological function of these sequences, and which belie the suggestion from Sequence Logos (Figure 1D) that position 2 contains no information.

[Figure 4 about here.]
[Figure 5 about here.]

To put these needs and features in a biological context, the data shown in Figure 1 are gene sequences belonging to a subset of Archaeal transfer RNAs, and are the binding motif for an endonuclease that removes an intron[2]. The core of the biologically relevant motif is shown in Figure 4, with the positions numbered as shown in Figures 1–3. The paired nucleotides on the opposite sides of the upper stem, internal helix, and lower stem regions each mutually influence each others' identities though well-known Watson-Crick nucleotide interactions. Regardless of the statistical strength, or magnitude of the edges found between these in the training data, a biological end-user would prefer to retain these dependencies in the model, because proper Watson-Crick pairing is essential for this motif's biological function. In addition to these predictable dependencies however, there are additional interactions present between several unpaired positions, particularly in the 5' loop. If we manually wrap the dependency structure shown in Figure 3 around the biological structure, we arrive at Figure 5. Despite the fact that the majority of the interactions present are not between sequential neighbors, it is critical to the biologist studying such a system, that the ordered, sequential property of the nucleotides is maintained in any representation. It is also critical to represent dependencies not just between the positions/nodes, but between the observed categories within the nodes, even for sequentially distant positions. At the same time, to support the researcher trying to model their data, none of the interactions can be arbitrarily sacrificed for simplification or clarity without inspection.

For practical applications, the researcher needs to be able to address similar problems with hundreds of sequential positions, and dozens of possible categories, and for which there is no simple physical structure to guide the layout. This makes manual layout and edge-routing impractical as a general approach.

*Overridingly, while all of these needs could be addressed in, for example, a "small multiples" fashion by something as simple as graphically-represented contingency tables, a medium-sized sequence family with 300 positions, would require visualizing $\binom{300}{2} = 44850$ contingency tables. Visually integrating these to develop*

*an understanding of patterns in the data quickly fails to inattention and change-blindness issues, and so ideally the end user needs all of this data to be presented seamlessly within a single visualization.*

Results of Biological Application

We have applied these ideas in the development of a prototype visualization system, StickWRLD, and used this system in collaboration with several biological labs to create novel and powerful statistical models that are being used for productive work today. While StickWRLD was originally developed as an expedient solution to visually explore evolutionary dependencies in biological sequence families, our recent work has converted this system from one which simply displays dependencies, into one that supports the development of complex predictive statistical models for the dependencies it displays. And, as reported here, these statistical models are superior to models developed without an adequate understanding of the interdependency structure of the model features.

Amongst these are projects that examine the protein sequence–function relationship, and that identify nucleic-acid sequence motifs that are intractable to traditional alignment and search methods due to interaction of both sequence and structure information. In addition to these end-user projects that we briefly report on here, StickWRLD has also been applied to identifying interactions between treatment variables and their concerted effect on outcomes in premature infant care(in press), expression Quantitative Trait Locus analysis[3], and Personalized/Precision medicine[3]. In all of these applications, the complex statistical models that have been successful, would not have been possible without a tool that supported visualizing and exploring the complex networks of conditionally-interacting features that are present in the data. While we do not propose that StickWRLD is an ideal solution for visualizing these features, we suggest that it is a prototype for building these important models, from which improved tools may be derived.

Through the rest of this manuscript we will: examine the properties of a particular variety of GPM, the Conditional Random Field(CRF), that make them particularly appropriate for modeling many types of biological data and that must be usefully conveyed in visualizations for them; highlight the utility of CRFs in 2 distinct biological applications; illustrate the representational needs of CRFs and their similarity to categorical parallel coordinates; and suggest extensions to the parallel coordinates paradigm that we have found useful for applying CRFs to biological-domain problems in our group and for our collaborators.

Graphical Probabilistic Models

GPMs have a long and convergent history, originating in several fields including physics[4], genetics[5] and statistics[6, 7]. In each, the idea originated as a means to describe the interaction of variables. The common paradigm is of a set of nodes that describe variables or marginal distributions of variables, and a set of edges that connect these nodes, which encode the joint distribution of variables in the nodes that they connect.

Amongst the simplest GPMs, Markov Chain models are an example of a chain-topology probabilistic graphical model where the training data is used to generate a sequence of states, and transition probabilities between sequentially neighboring states[6, 8]. While such a model is typically thought of as generative, it can be used to determine the probability that a sequence of observed data was generated by the same process that produced the training data, essentially by walking the chain of states, following transitions based on the observations. Applications of this nature are frequently found in bioinformatic questions such as "is this gene a member of the same family as the genes in my training set?" Markov Chain models however, are memoryless. That is, the conditional distribution of future states in the chain at any given state, depends only on the current state, and not on the series of states that preceded it. Therefore, the transition followed based on an observation, depends only on the current state and the observation. This limitation is appropriate, only if the underlying data domain truly obeys this memoryless "Markov Property". If the underlying data can contain dependencies on distant states, violating the Markov Property, Markov Chain models are at best approximations of the characteristics of the training data.

*Significantly, Markov Chain models can be well-represented by Parallel Coordinates visualizations of the node and transition structure. This near isomorphism (the potential for Markov Chain states to loop is omitted) between Parallel Coordinates and association rules on item sequences has been previously reported by Yang[9]. We propose that there is a more complete isomorphism between some classes of Graphical Probabilistic Models, and parallel axes on which a fully-connected graph for each feature vector is projected. The fundamental mappings are between nodes and categorical parallel axes, and conditional weighted edges and linked categories on the axes. Limitations on this mapping, and potential extensions to the parallel axis schema to overcome these limitations is outside the scope of this manuscript, but is the subject of another manuscript in preparation.*

Generalized Graphical Probabilistic Models attempt to overcome the limitations of chain models, at least conceptually, by encoding arbitrarily complex networks of dependencies between states. For classification purposes, this provides significant benefits over previous methods that were limited to either assumptions of strict independence between features, or, assumptions of Markov Property memoryless dependence. Again conceptually, this means that GPMs can encode models for domains that violate the Markov Property. Such problem domains abound in areas from the biological sciences, where protein function is modulated by the dense network of contacts between amino acids in a three-dimensional structure, to economics, where stock prices are influenced by a dense network of suppliers, consumers and competitors. From identifying sites in the genome that possess complex combinations of signal sequences, to linguistics, to medical diagnoses, where a problem domain possesses interaction networks more complicated than linear graphs, GPMs that can encode this additional network information, produce more accurate results than linear chain models.

Unfortunately, these features are largely conceptual benefits of generalized GPMs, because due to violation of the Markov Property, network-connected GPMs cannot be "stepped through" in the same fashion that Markov Property chains can. Instead, to evaluate an observation at a particular state, the observations at all states connected to that state must be evaluated. If the graph-connectivity of the GPM is such that it contains cycles, then all of the nodes in the cycle must be evaluated simultaneously. As a result, the successes of GPMs, to date, are limited to domains where the interaction network is tree structured, or, where there are few conflicting observations found along any cycles in the network. For complex connectivity with many overlapping cycles and biologically realistic noisy data, annealing optimal node and edge weights to correctly represent the training data, becomes computationally intractable.

As a result, GPMs have the peculiar property that it is quite easy to describe the "conceptual GPM" that models a collection of training data – one simply builds a graph with nodes for the measurable features in the training data and connects them with edges describing the dependencies – but it is quite hard to convert this conceptual model into one that is actually computable. The currently extant solutions involve heuristic unrolling of cycles, or manual specification of the dependency graph, limiting practical applications of GPMs to either quite small, or to topologically simple problems.

For the purpose of this paper we are interested in undirected GPMs in which:

- The dependency network may (theoretically) be complete across the nodes.
- The node weights are conditional on the observations.
- The dependency network edges and weights can be conditioned on the observations.

Such models, where the set of variables over which a joint distribution must be considered, and the weightings of their combinations are dependent on the values observed for the variables, are exemplified by Conditional Random Fields (CRFs). Much of the work presented here is guided by requirements for working with CRFs, but it is equally applicable to simpler densely connected dependency graphs models as well.

CRFs were originally described by Lafferty et al. in 2001, as an alternative to Markov Chain, and other GPMs, for building probabilistic models to segment and label sequence data[10]. Their development was motivated by the inability of Markov Chain models to address multiple interacting features and long-range dependencies between observations, and by branching biases in other models.

Lafferty proposed that CRFs be constructed by explicit manual specification of the connection topology for the states, and heuristic determination of transition parameters to fit the training data characteristics onto this *a priori* specified topology. In their development, Lafferty considered fully-hierarchicalized linearizations of the training model, and rejected these due to the potential combinatorial explosion that can occur if the training data implies many dependent transitions at each state (effectively, fully-hierarchical linearizations quickly reach a state where the number of linearized sub-models exceeds any possible number of observations in the training data, resulting in dramatic over-fitting errors). Likewise, initializing fully-connected training models and annealing them into a tractable state was also considered, and rejected due to difficulties in imposing prior structural knowledge on the final model.

Lafferty demonstrated that CRFs with low-order models of higher-order data, outperform chain GPMs with equivalent limitations, however the exact relationship between the predictive accuracy of a CRF model, and the detail with which it reproduces the real dependency structure of the training data, has not, to our knowledge, been described.

Several schemes have been proposed for using training data to estimate parameters in a computably-simple CRF, including two in the original description, and others that attempt to enhance the accuracy of CRFs for data that contains higher-order dependencies in the actual data distribution, than are encoded in the model. None of these produce stable solutions for systems containing complex graph connectivity, and usually only perform well with topologies no more complex than isolated cycles with no shared nodes or edges.

## Results and Discussion

The results we present here are the culmination of several years' analysis of what is required to solve typical biophysical domain tasks using GPMs in our labs, and those of our collaborators, as well as two examples of problems to which we have applied these techniques, through an experimental approach to meeting the analysis needs.

### Analysis of Typical Domain Tasks

A typical end-user comes to the world of GPMs with a collection of training data, and a desire to use that data to build a model that can identify other data that are "like" the members of the training set. It is trivially easy to develop a model that accepts only data that is *identical* to members of the training set, but developing one that accepts things that are *similar* can require considerably more insight into the important features of the data, and into exactly what is meant by "similar". Traditionally, if the important features are not either statistically independent, or the user cannot *a priori* define the important dependencies, the standard best practice has been to feed the data to a chain-model GPM such as a Hidden Markov Model, and to hope that whatever other dependencies exist, they aren't such critical features as to make the chain-model GPM completely irrelevant.

*We note, with some foreshadowing, that this situation is strikingly analogous to users relying on traditional parallel coordinates' representations of the correlations between "sequential" axes, as a hopeful proxy for the full complexity of the data.*

The overriding goal towards which our work is therefore directed, is displaying and facilitating user-interaction and editing of the complete data-implied dependency structure from a collection of data, into a computationally tractable dependency structure from which existing GPM parameter estimation methods can model the training data.

**Visual Features** For the bio/life-scientist then, in addition to meeting the **User Requirements** the important features of a visualization or visual analytics tool directed at developing CRF models are:

1. There must be a "node" concept that maps to the biologist's understanding of a measurable feature in the data.
2. The natural ordering of the nodes should be maintained to provide context and landmarks.
3. The possible categorical values of each node must have distinct representations (we consider these to be subnodes).
4. Identical categories (subnodes) in different nodes, must have identical representations.
5. The subnode representation must be able to encode the weight of each category in each node – for example, the marginal probability of observing each category at that node.
6. There must be an "edge" concept that maps to the biologist's understanding of a relationship between different measurable features in the data.
7. The edge representation must be able to convey "when we observe category $M$ at node $j$, we will observe category $N$ at node $k$".
8. The edge representation must convey the strength of the expectation – for example the joint probability of observing both linked categories.
9. The edge representation must be able to convey the extent to which the joint probability is predictable from the marginal probabilities – i.e., the extent to which the features are dependent or independent.
10. The edge representation must be able to simultaneously convey such relationships between $M$ at $j$, and any other categories, at any other nodes.
11. It must be possible to simultaneously visualize all relationships for $M$ at $j$, as well as all relationships for any other category found at $j$, to all of their respective subnode partners.
12. It must be possible to simultaneously visualize the complete set of relationships between subnodes, for every node in the data.
13. The representation must be robust to simplifications – the node landscape and sub-node arrangement must be invariant to filtering or changing subsets of displayed joint-probability edges.
14. No edge may be occluded within any other edge or edges (the layout of the subnodes must be such that no subnode-to-subnode edge is co-linear with, and overlaps any other edge).

Identification of these **User Requirements** and **Visual Features** are the accumulated work of almost 20 years of continuous application in our own biophysical work and in collaboration with several other biological labs, and the evolution of a system to address the complex system modeling needs that we have discovered that our diverse domains share. The results of addressing these requirements and providing the desired features, has been data and insights, unaccessible by other means, of sufficient interest to have resulted in numerous papers, and funded research projects for our, and several other labs.

### Approaches using Existing Parallel Coordinates Representations

Parallel coordinates[11] are an interesting visualization paradigm to consider in relationship to GPMs, because they can, with minor adaptation, be used to visualize and manipulate the dependency structure of the chain-structured subclass of GPMs.

In a traditional parallel-coordinates plot, the multiple axes of a high dimensional space are arrayed in parallel on a plane, rather than being arranged orthogonally. Each multidimensional feature vector in the data, is then displayed as a polyline that passes through each parallel axis, and that links the coordinate that it possesses on each axis. Parallel coordinates plots have interesting applications in computational geometry, because points lying on or near certain high-dimensional geometric surfaces produce distinct patterns on the parallel axes, enabling the presence of these surfaces to be detected visually.

More frequently however, parallel coordinates are used to visualize and understand general patterns of dependencies within high dimensional data. Parallel coordinates provides advantages for this use, over visualizations using isometric-style 2-dimensional projections of high dimensional spaces because it does not ambiguously collapse an entire line of high-dimensional points into a single displayed 2D point – every point in the high-dimensional space can be unambiguously represented in the parallel coordinates plot by a distinct polyline. In exchange for this increase in representational accuracy, parallel coordinates trade complexity with respect to the ordering of the axes – changing the ordering in which the parallel axes are shown, can generate dramatically different understandings of the patterns in the data.

Parallel coordinates have typically been applied to continuous-valued data, and using strictly parallel representations of the axes, but more recent developments have explored extensions that broaden both the approach and the application. Approaches for categorical data[12, 13], multiply-connected axes[14], and arbitrarily arranged axis segments in the plane[15] have been developed. To successfully connect many-to-many parallel coordinates axes in a planar plot, Lind[14] and Claessen[15] both sacrifice polyline connectedness for their point representations, and accept that a single axis may have multiple representations in the plane, to capture all necessary disjoint polyline edges. The work on many-to-many, multiply connected, and flexible linked axes attempts to address the issue that the information transfer from parallel coordinates plots, depends on the axis ordering. Other work has attempted to address this same problem by identifying the "best" order for the axes, based on statistical measures of shared information between the axes[16, 17, and many others]. Most of these developments have focused on retaining the traditional restriction of parallel coordinates to axes in a single plane, to avoid 3D occlusion and view dependency issues.

In addition to 2D representations, extensions of parallel coordinates into 3 dimensions have been approached in several ways. Fanea[18], Johansson[19] and Kerren[20] have each attempted very different representations. Fanea's approach uses 3D to decompress the overlaid polylines representing each data point in traditional parallel coordinates, by "fanning out" each axis into multiple representations of itself, each slightly rotated around the horizontal axis. This representation improves understanding for the behaviors of individual polylines in the image, and in a small-multiples sense, enables the user to mine more distant relationships than traditional parallel coordinates, by visually tracing individual polylines. Johansson's approach moves still-parallel axes off the plane and into an axially-aligned star topology around a central axis, and displays relationships from each of the "tip" axes back to the singular central axis. This approach enables an enhanced view of a single coordinate, but sacrifices connections between the star tips to accomplish this. Kerren's 3D Kiviat diagrams are a hybrid of categorical parallel coordinates, and Fanea's "fanned out" axes. In the 3D Kiviat diagram, categorical parallel axes are laid out radially from a single point forming a sort of parallel-coordinate star diagram. Individual points of the star can then be interactively "fanned out" into additional representations of that axis so that details of the poly-category-line trajectory can be more easily seen.

While all of these parallel coordinates approaches enable some intuition into a subset of relationships amongst the axes, none attempt to display the entire relationship structure simultaneously. Moreover, the primary intent of the traditional parallel coordinates approach, and of its many derivatives, has been to represent the features of the data, rather than to provide an interface by which the important features could be selected and combined into statistical models. As a result, none are well-tuned to the task of extracting tractable GPMs from interaction patterns in visualized data.

Despite these limitations, if one re-imagines parallel coordinates to be visualizing a multigraph/metagraph as described previously, categorical parallel coordinates becomes a good reproduction of a chain-structured linear GPM. Each axis becomes a node, with subnodes arrayed along its length. The relative extent of each category on an axis, corresponds to its relative sub-node weight in the GPM. Likewise, the magnitude of the edge between categories corresponds to the "transition" weight accumulated by the GPM when shown a feature vector that contains that particular pair of subnodes. This re-imagining can be most immediately applied to linear Markov Chain models. A Markov Chain model can be thought of as representing the characteristics of a collection of time-series data. The model presents each point in time as a collection of possible states. We can consider each time-point to be a node. From each state "in the present node", edges describing transition probabilities pass to the states at the next time point/node "in the future". Any particular time-series in the training data, describes a trajectory through a specific state at each time point/node. If we map each time point/node to parallel-coordinates dimensions, then each time series in the training data describes a multidimensional vector with state-values as components. This structure naturally fits a parallel coordinates representation, with categorical-valued axes.

Unfortunately, while traditional parallel coordinates capture the features of a linear-chain GPM (where only interdependencies to the immediately preceding and following nodes must be understood), they cannot reasonably accommodate the arbitrary dependency structure of a random-field GPM such as a CRF. If categories are assigned ordinal values, traditional parallel coordinates can satisfy **User Requirement** 1, and partially accommodates **User Requirements** 3 and 5. It also addresses **Visual Features** 1-4, 6, 7, and 13.

Parallel Sets[13], is more recent evolution of parallel coordinates, better adapted to categorical data. Parallel sets have features that convey subnode weights and edge weights, improving information transfer for these features. Moving to Parallel Sets again satisfies **User Requirement** 1, additionally satisfies 2, and further addresses 3 and 5. Amongst the **Visual Features**, it retains 1-3, 6, and 7, sacrifices 4 (Parallel Sets also sacrifices 13, but this is an implementation issue), and additionally satisfies 5 and 8.

While both traditional parallel coordinates, and categorical parallel sets representations show a plethora of interesting features about the data, neither are particularly informative regarding the complexly connected, biologically important patterns in the data, and neither provide intuition about how to build a biologically relevant, computationally useful model of the training data. In particular, our domain knowledge says that there are likely to be dependencies between sequentially-distant features, due, for example, to folding of molecules. These sequentially-distant dependencies are often more important signatures of family membership, than either the specific positional identities, or sequentially-proximal dependencies. To address this, at a minimum **User Requirements** 4 and 5, and **Visual Features** 10-12 must be also be satisfied. Additionally satisfying **Visual Features** 9 and 14 are a significant priority.

## 0.1 Extending Parallel Coordinates to Address Additional Requirements

What is striking is that the manually laid-out dependency and weight diagram shown in Figure 5, is not very different from a Parallel Sets representation, but now satisfies **User Requirements** 1-5, and **Visual Features** 1-7, and 10-13. Capturing **Visual Features** 8 and 9, requires only adding visual weights to the edges.

The primary representational differences (aside from a layout conformed to the shape of Figure 4 for convenience), are that Figure 5 does not waste ink redundantly displaying joint distributions that are entirely implied by the marginal distributions, and that it does not restrict the displayed edges to only being between sequential nodes.

*This linkage of non-sequential nodes, is critically important to adequately understanding the data because the data cannot be assumed to obey the Markov property. In fact, understanding where, and to what extent the data violate the Markov property, is a defining characteristic of the biological needs in all of the domains that we are trying to address. No representation that limits the display of dependencies to "sequential" positions, will ever inform the user of non-sequential dependencies.*

We know about some of the dependencies in the Archaeal tRNA splice sites *a priori* from domain knowledge, but importantly, given an adequate visual representation, the data can inform us of these dependencies even without that prior knowledge. It can also tell us about other dependencies that are no less important signatures of family membership, but for which the domain knowledge is silent.

While similar biological data can only rarely be attached to a biologically-relevant shape, an expedient, automatable, and near-universally applicable general solution which simultaneously satisfies **Visual Feature** 14, is simply to allow the parallel-coordinate axes off the plane, and array them around a cylinder.

[Figure 6 about here.]

With categorical parallel-coordinates axes arrayed around a cylinder, and fixed categories arrayed at specific locations along the cylinder's length, we can overcome the Markov-Property-like character of the polyline used to represent each feature vector in traditional planar parallel coordinates, and replace this polyline with a formally complete undirected graph between all of the subnodes traversed by the feature vector. If we cast a set of feature vectors into this space, and weight the subnode-to-subnode edges based on the number of features sharing those sub-nodes, we can visualize the entire $\binom{SequenceLength}{2}$ set of contingency table joint and marginal distributions in the same figure. Figure 6 shows the results of this approach. It is clearly cluttered, and on paper, densely occluded and difficult to interpret, but even with these impediments it is already showing us that there are quite strong patterns of co-occurence between the **A** at position 2, with **C** at 1, **T** at 3, **A** at 4, and also with occluded sub-nodes at several other positions. The strength of the $A_2 \leftrightarrow C_1$ and $A_2 \leftrightarrow T_3$ relationships are visible with some study of Figure 6, however the other relationships with $A_2$ are not conveyed by any canonical alternatives. Because these dependencies involve non-sequential columns, even with this limited view, this intuition is beyond what is easily attainable with traditional parallel coordinates or parallel sets.

If we further calculate the difference between the observed joint distributions, and the predicted joint distributions based on the marginal distributions, and use these as edge weights instead of using the observed joint distributions, we can eliminate ink wasted on joint distributions that are entirely predictable, and focus the user's attention on the patterns of dependencies they need to understand.

*The Requirement for Interactive Analysis*
One step remains to convert this visualization both into something visually understandable, and simultaneously into a dependency structure amenable to creating a computable CRF; engaging the user in the task of simplifying the dependency structure. The raw dependency structure implied by the training data is often both too complex, and too specific for practical use without further refinement. The primary interaction required is for the user to filter the dependencies displayed, such that those that are reasonable based on domain expertise remain, while as many others as possible are eliminated.

In addition, we have found that for some tasks, a simple threshold is insufficient to segregate the important and unimportant dependencies. In some cases to capture biologically important features of the model, it is necessary to let the user retain dependencies with statistically or quantitatively small weights. A prototypical example of this need is demonstrated by stem structures in nucleic-acid sequence families. In these cases a domain user knows that there are specific nucleotide-to-nucleotide pairings allowed, and may know that certain positions in the sequence absolutely must be paired, to retain functionality. It does not matter if the training data represents all of the biophysically relevant pairings with similar frequency for those positions – the user must be able to retain the absolute conditional dependency edges for the paired positions, even if some combinations only appear in the training data with very low frequency. In other cases, users prioritize retention of coherent groups of weaker dependencies over scattered weak dependencies, and sometimes even over scattered stronger dependencies, based on domain intuition regarding the biological origin and function of the dependencies.

The choices involved in selecting these groups are partially influenced by domain knowledge, partly by a learned understanding of certain archetypal visual motifs that appear in the domain data, and partially by untrained visual intuition. The exact mechanisms applied, and how to best support them in a user interface remain to be studied in greater detail, however, it is clear that this interactive selection process provides a mechanism for exploratory experimentation with the structure of the CRF, wherein users can easily try different choices for retaining or excluding dependencies.

[Figure 7 about here.]

To support this final interactive refinement of the raw dependency structure into a tractable subset, our experimental StickWRLD interface enables the user to adjust the residual magnitude and significance thresholds (and several other threshold parameters) for selecting the subset of the raw dependencies to display. We also detect edge cycles and highlight these for the user with edge coloring, and enable the interactive selection and removal of edges from the dependency data structure. We are currently experimenting with on-the-fly parameter estimation for several GPM varieties and estimation algorithms[21] and painting of the visualized edges with the estimated parameters. If acceptable performance can be attained, this shows promise for informing the user of situations where a removed edge dramatically affects the GPM parameter landscape.

Putting all these things together, we can simplify the parallel coordinates in a volume view, starting from where we began in Figure 6, and ending in a visualization such as Figure 7:C. As with all representations, the strong dependencies between **G** in column 7 and **C** in column 8, and vice versa are apparent. The similar strong dependence between $G_6$ and $C_9$, and $C_6$ and $G_9$ is now also visible. Several unexpected dependencies have also appeared amongst columns 1, 2, 3, and 4. This simplified model of the Archaeal tRNA sequence motif makes surprisingly good predictions about other candidate sequences' biological functionality, when these alternative sequences are biologically substituted in place of members of the training set[2].

*Alternatives, and Issues with 3D layout:*
Extending traditional parallel coordinates to higher dimensions, effectively displaying the complete graph of each element of the training data, upon the parallel axes in the plane addresses the concern that dependencies between sequentially distant columns are invisible in the traditional parallel coordinates representation, but simultaneously brings to the fore a host of complications ranging from the fact edges can now be co-linear, obscuring their actual positions, to the potentially overwhelming clutter that appears when visualizing *every* edge of every element in this fashion. In fact, with even a small amount of heterogeneity in the identities found in the training data, a complete-graph-per-feature visualization in a planar figure quickly devolves into a completely uninformative image where every possible edge is displayed, and there is no visual weight given to any of the important features. Attempting to alleviate these difficulties by, for example brushing and linking from a secondary display of the training data provides some improvements, but relies on the user's memory to identify clusters and commonalities in the trajectories of the training data through the axes. A similar extension of parallel sets meets with similar difficulties, and even greater visual clutter.

Claessen attempted to deal with connecting each parallel-coordinate axis to more than two neighboring axes, by giving each axis multiple representations in a planar figure[15]. This paradigm may be useful for representing biological data of this nature for sequences with quite limited length, or for re-representing subsequences of data from longer sequences. However, for surveying the dependency structure of large sequences, this approach fails to the same issue that prevents small-multiples contingency tables from being useful. The need to maintain natural ordering aside, fundamentally the researcher cannot know which pairs of axes are important to look at together, until they have looked at all of them together. A typical sequence family of length 300, which has 300 actual axes amongst which dependencies must be explored, would require over 22,000 displayed planar copies of these axes (half as many as the contingency tables, since each visualized axis can display dependencies with two neighbors rather than one). Keeping track of which are replicates, and traversing complex networks of dependencies within this display would not be practical.

Wrapping categorical parallel coordinate axes around a cylinder clearly violates the visualization design rubric that good visualizations should constrain themselves to 2-dimensional representations. It also obviously introduces issues where occlusion is viewpoint dependent. However, even our simple 9-column example would require 36 purely 2D plots to present the data without obscuring edges, and because we are interested in combinations of ($\geq 2$) columns with dependencies, we would need to look at *every possible ordering* of those 36 plots. In practice, applied over almost 20 years to real data from several collaborating labs, the impediments induced by the 3D visualization are overwhelmingly outweighed by the benefits of being able to see all of the data in a single interactive model. Application of these techniques, in our lab and those of our collaborators, has regularly been found to replace months of laborious examination of 2D contingency results, with minutes of interactive exploration of a 3D model. This benefit accrues even when the competing 2D contingency tables are supplemented with a planar node-to-node dependency graph as an index into the sub-node contingencies.

Despite these successes, we do not claim that this representation is optimal, and many possible alternatives remain to be explored. Chief amongst these are interactive techniques where a planarized subset of interdependent features is displayed in a brushing-and-linking fashion based on selections in a circular node-to-node (rather than subnode-to-subnode) overview. We argue only that our results unequivocally demonstrate that conveying full, weighted networks of subnode-to-subnode dependencies, is a critically underserved need in many biological domains, and that approaches to conveying this information to the end user provide enormous analytical benefits. Our hope is that other researchers will identify alternative representations that maintain the analytical power of our 3D presentation, while eliminating its less desirable side-effects.

[Table 1 about here.]

## Case Study : Protein Mutations and Function

Adenylate Kinase (ADK) is an extensively studied and characterized enzyme with a unique molecular/sequence feature[22, 23, 24]. Across evolutionary history, the family of ADKs has bifurcated into two groups that have the same protein structure, but that produce this structure using quite different biophysical stabilizing forces, produced by quite different amino acid residues in each family. Most prominently, one subfamily possesses a tetra-Cysteine Zinc-chelating motif, while the other stabilizes the same structure using a hydrogen-bonding network between His, Ser, Asp and Tyr in the same locations. The latter four are also associated with the presence an Arg and Glu in nearby positions, while the tetra-Cys motif is ambivalent about these positions[25]. This has made ADK a popular protein in which to study the relationship between protein sequence and protein function. Because each subfamily has an almost equal number of members, naïve models that look only at the residue distribution in the family, suggest that substituting any residue from the hydrogen-bonding subfamily into the tetra-Cys subfamily, should have no effect on function. Not surprisingly, this turns out not to be the case[26]. The acceptability of such residue substitutions is conditional on the context in which they are put. In fact, even swapping the complete hydrogen-bonding tetrad for the tetra-Cys motif, still results in a non-functional protein.

[Figure 8 about here.]

Following our earlier work in which we described a more extensive network of ancillary dependent residues around both the tetra-Cys and hydrogen-bonding networks[25], we developed a CRF that accurately predicts the changes in ADK function (enzymatic activity) that are produced by multi-point mutations in its sequence. The ADK family was visualized using StickWRLD, and the dependency network found in it iteratively refined to select the 4, 6, and 12 most strongly interdependent residues. Several steps in the refinement process are shown in Figure 8. Figure 8:A begins with the refinement already well under way, with the initial roughly 4 million edge raw dependency structure reduced down to several hundred edges using residual threshold cutoffs. Figures 8:B-D show additional refinement using statistical cutoffs for the remaining residuals, with Figure 8:D passing beyond the optimal refinement and losing significant portions of the dependency network due to too-stringent filters. After eliminating the majority of the positions with only minor dependencies, we settled on Figure 9 as the core of the CRF from which to select our 4, 6, and 12-dependency networks. CRFs were defined using each of these dependency subsets. Several varieties of ADK mutants were also made, with an assortment of substitutions from the hydrogen bonding subfamily, into *Bacillus subtilis*, which natively possesses a variant of the tetra-Cys motif.

[Figure 9 about here.]

To evaluate these predictions, we constructed mutants of *B. subtilis* ADK. *B. subtilis* ADK contains a rare variant of the lid that uses three Cysteines and one Aspartic Acid. The mutants were *B. subtilis* domain substituted with: the four hydrophilic residues (**Tetra**); the two associated residues (**Di**); and all six hydrophilic-motif residues (**Hexa**). A chimeric mutant (**Chim**) containing two of the Cysteines and two of the hydrophobic residues is known to be non functional[26]. Structural stability and enzymatic activity were assayed for each mutant. The wild-type *B. subtilis* sequence, and each of the mutants, were also scored by each CRF. Table 1 shows mutations created, the results of the biological assays, and the score produced by CRFs using 4, 6 and 12 nodes of the relationship network visualized in Figure 9. As expected, the mutant activity correlated directly with the extent to which the residues identified by the largest CRF, were replaced in the *B. subtilis* background. This alone is a significant finding in the protein sequence/function domain (Callahan, Perera, Weppler, Ray, Magliery, manuscript in preparation). Moreover, not only did the visually-refined CRF accurately identify the residues that were necessary to swap to transfer function, *the most complete CRF also accurately predicted the extent of functional loss, in each of the mutants*. While still requiring further research and validation, it appears that the 12-dependency CRF's scoring of "this sequence is a good match for the training data" correlates with the stochastic probability of the enzyme's catalytic reaction taking place. In other words a mutant sequence that scores similarly to members of the training data, will have activity like the members of the training data, while mutant sequences with scores significantly different from the training data will have activities that differ, in correlation to the differences of their scores.

There are a number of algorithms that use statistics based on per-position residue frequency to predict the functional consequences of mutation[27, 28, are amongst the most popular]. None of them can make accurate predictions in this protein, because the consequences of a mutation depend on other residues in this protein, not just on the mutation itself. Even HMM-based methods that evaluate sequentially-proximal dependencies are unable to accurately predict these functional changes, because the dependency network is both dense, and spans over 50 positions.

Only the CRF model is able to make accurate predictions regarding functional changes. Notably the predictive correlation does not appear for the 4-dependency CRF, begins to show correlation with the 6-dependency version, and does not become completely predictive until 12 dependencies around the primary (Cys or hydrogen-bonding) tetrad are included in the CRF. This not only supports our contention that more complete GPMs make better predictors, but also highlights the importance of simultaneously visualizing the more complete dependency graph over the categorical parallel coordinates. The 12 most predictive residues were identified because of their complex and highly connected dependencies with the well-documented primary tetrad - not based on the statistical strength of those dependencies.

[Figure 10 about here.]
[Figure 11 about here.]

Case Study : Polyadenylation signaling DNA motifs

A completely different problem is presented by the question of identifying the genomic signals that govern the addition of the "poly-A" tail to messenger RNA molecules. Messenger RNAs are molecules that are used to transmit the genomic blueprint for proteins from an organism's DNA, to the cellular machinery that makes proteins. The longevity of these messages, as well as several other features of their use by the cell, is governed by the length and location of poly-adenosine-monophosphate tails that are independently added to the message after it has been synthesized. The signals that direct this polyadenylation are not well understood, and modulating polyadenylation is an interesting research focus with potential impacts that range from fighting disease to biofuels. We have been developing improved models of two different types of polyadenylation signals found in the human genome.

[Figure 12 about here.]

Unlike with Adenylate Kinase, where an unambiguous alignment of the proteins lets us say "the symbols in column $i$ of each data vector all are functionally equivalent", in the case of polyadenylation signals, we know neither the pattern, nor exactly how the sequences should be aligned. What looks like column $i$ in one member of the training data, can be column $j$ in another. In general, these shifts are small, but they result in either weakening of the apparent specificity of the model, or in the generation of an unnecessarily complex model that contains separate sub-models to address each of the alignment possibilities.

Rather than accept either of these non-optimal modeling situations, with this data our task is not simply to model the training data, but to successively refine a model derived from some of the training data, such that more of the training data fits, and improves the model, with each refinement. The ability to visualize the dependency structure, and to interact with and edit the dependencies to generate a model, is once again critical for developing an accurate understanding of the sequence family properties, and to generating an accurate model that can select and align polyadenylation signals correctly.

The starting point for this analysis was a pair of data sets, both derived from genomic regions purported to signal for polyadenylation. The first data set contained sequences in which traditional sequence-similarity metrics such as Position Specific Scoring Matrices (PSSM) and Hidden Markov Models (HMMs) had detected a consistent pattern. The second data set contained sequences for which PSSM and HMM methods failed to find any pattern, and in which the models created using the first data set, failed to identify matching regions. Visualizing the first, "signal" data set, as shown in Figure 10, we see a strong pattern in the marginal distribution of bases in each column (which is what the PSSM and HMM methods identified).

Our real focus in this project however, was in identifying any signal in the data where there was thought to be none. Visualizing this "non-signal" data, as shown in Figure 11, it is clear that distribution of bases, at least with the original alignment of the sequences, is uninformative. However, within the dependencies, we see a peculiar feature: There is a pattern of dependencies between bases in several positions, that appear as a repeated "echo" of the same dependency between the same categories, shifted to different columns. This repeating dependency pattern is a sign that some of the sequences are misaligned. By interactively selecting the sequences that participate in the misaligned echo (a feature available through the StickWRLD interface) and aligning the *dependencies* (rather than the bases) within the data, we are able to correct the alignment of the "non signal" data to the state shown in Figure 12. This new model of the "non-signal" signals demonstrates that these signaling regions actually do have a strongly conserved regulatory motif that is not very different from the previously well-defined "signal" motifs. It allows somewhat more variation in base identities than the "signal" motif, and is dominated by different dependencies between positions and bases, but it is nonetheless a distinct identifiable pattern. Biological validation of this new "non-signal" model is ongoing in our labs.

While this result itself is quite important, the real significance of this work is that we have successfully aligned a family of sequences that couldn't be properly aligned based on the per-position sequence statistics, by using the discovered and visualized dependency structure found within the data.

## Conclusion

Moving parallel coordinates from the plane, into a volume in an interactive interface, enables the complete-graph nature of dependency networks to be visualized, understood and used in a fashion that is not possible with the pairwise dependency information to which 2D representations restrict the analysis. The ability to do this is critical for improving the utility of GPMs, which show significant promise for modeling many types of data where features interact, but which have thus far seen only restricted applications because of the difficulty in developing tractable, yet appropriately connected dependency networks for complex data.

We have demonstrated the application of a modified form of parallel coordinates in a volume, using our prototype implementation StickWRLD, for two very different types of GPM modeling problems. The first involved the development of a CRF where the evaluation "this new sequence is like (or unlike) the training data" can be used to predict that the functionality of a changed sequence will be like, or unlike the functionality of the parent. The second involved iterative refinement of a GPM from poorly aligned training data, where the detection of similar, but offset dependency networks in the training data are used to realign the training data, repeatedly strengthening the model. In both cases the predictions of the GPMs are being validated in our labs, with manuscripts in preparation for each.

By applying appropriate visual weight to edges, and eliminating from view, everything but the most important features, the volume version of categorical parallel coordinates can be transformed from a hopelessly complex representation, into a useful visual analytics tool in which users can explore the effect of different parameter choices and interactively select the data-implied dependencies to incorporate into a GPM model. Numerous filtering and selection schemes, as well as domain-appropriate feature clustering and display simplification schemes are present in the StickWRLD software interface, and the fundamental representation paradigm can be adapted to any form of mathematically symmetric statistical association detection. It is clear from testing that surprisingly simple statistics provide useful insight for building GPM models – simple residuals, as we originally described when casting StickWRLD as simply an exploratory tool[29] remain one of our most common approaches – though more sophisticated analyses ranging from Fisher's Exact Test of Association[30] to data input from arbitrary external applications have been applied as well. Experience with these options and experiments performed with them[31] suggest that there is no single scheme that is ideal for all data analyses, further underscoring the need for robust exploratory interfaces that enable the user to see and explore the network of interactions revealed by different approaches and focused upon by different filtering schemes. Experiments and optimizations in this area are ongoing.

Our successes with parallel coordinates in a volume demonstrate the improved understanding that derives from more complete dependency visualization, but they simultaneously open the door to new questions about the most appropriate statistical measures to aid the user in filtering the dependency network (is, for example, a simple threshold, the most appropriate way to segregate meaningful dependencies from unimportant ones?), and to new questions about the most appropriate representation for this type of data. Categorical parallel coordinates in a volume appear to be isomorphic to a type of metagraph in which one variety of nodes corresponds to parallel coordinate axes, with these nodes containing a second type of node, corresponding to ordered categories. Dependencies, as we have measured them, occur between the category-type sub-nodes, but there is no conceptual reason that dependencies between axes (for example, as detected by Mutual Information, or Joint Relative Entropy), or between categories and axes, could not also be incorporated in the model. Nor is there a reason that dependencies must be limited to pairwise interactions. Further exploration of visual paradigms for interacting with this more complete metagraph structure remain an interesting research topic, and will likely further improve GPM development tools in the future.

It is also likely that there are further optimizations of our visual representation that minimize the occlusion and complexity issues inherent in our 3D display. While our results demonstrate that static planar representations cannot provide the detailed insights necessary to address the domain needs for understanding complex metagraph-type-data, and that these needs are usefully addressed in the 3D domain, we do not suggest that there are not other interactive paradigms that might make these high-dimensional features accessible without requiring 3D exploration. We encourage further exploration of alternatives that retain the ability to display the full complexity of networks demonstrated by our StickWRLD experimental system.

## Acknowledgements
This project was funded partially by Nationwide Children's Hospital, The Ohio State University, and by an NIH-AREA award (1R15GM94732-1 A1 to CL)

**Availability of Supporting Data**
The data set supporting the results of this article related to Figure 1 is included within the article and its additional files.

**List of abbreviations used**
GPM: *Graphical Probabilistic Model*, CRF: *Conditional Random Field*, ADK: *Adenylate Kinase*, PSSM: *Position Specific Scoring Matrix*, HMM: *Hidden Markov Model*, RNA: *Ribonucleic Acid*, DNA: *Deoxyribonucleic Acid*, 2D: Two Dimensional. 3D: *Three Dimensional*

**Competing interests**
The authors declare that they have no competing interests.

**Authors' contributions**
WCR developed the visualization approach and identified the connection to GPMs, developed the experimental approach, wrote the prototype software versions and was the primary author of the manuscript. SW developed much of the release software and extended features, as well as ran test protocols and analyzed data. NC and TM conducted the experiments in ADK and analyzed the ADK data. MD, QQL and CL constructed the polyadenylation datasets and assisted in analyzing the polyadenylation data. CWB provided guidance and critical insight into the statistical properties of the visualization and helped connect the theoretical approach to research applications. All authors contributed to manuscript editing.

**Author details**
[1]Nationwide Children's Hospital, 575 Children's Crossroad, 43215, Columbus, OH, USA. [2]The Ohio State University, 100 W. 18th Ave, 43210, Columbus, OH, USA. [3]Miami University, 501 E. High St., 45056, Oxford, OH, USA.

**References**
1. Gaur, D., Shastri, A., Biswas, R.: Metagraph: A new model of data structure. In: Computer Science and Information Technology, 2008. ICCSIT '08. International Conference On, pp. 729–733 (2008)
2. Ray, W.C., Ozer, H.G., Armbruster, D.W., Daniels, C.J.: Beyond identity - when classical homology searching fails, why, and what you can do about it. In: Proceedings of the 4th Ohio Collaborative Conference on Bioinformatics, pp. 51–56 (2009)
3. Ray, W.C., Wolock, S.L., Li, N., Bartlett, C.W.: Stickwrld : Interactive visualization of massive parallel contingency data for personalized analysis to facilitate precision medicine. In: Proceedings of the 3rd Annual Workshop on Visual Analytics in Healthcare, in Conjunction with the American Medical Informatics Symposium. VAHC '13, pp. 68–71 (2013)
4. Gibbs, J.W.: Elementary Principles in Statistical Mechanics: Developed with Especial Reference to the Rational Foundations of Thermodynamics. Yale bicentennial publications. C. Scribner's sons, New York (1902)
5. Wright, S.: Correlation and Causation. J. Agric. Res. **20**, 557–585 (1921)
6. Markov, A.A.: Extension of the Law of Large Numbers to Dependent Quantities (in Russian). Izvestiya Fiziko-Matematicheskikh Obschestva Kazan University **15**, 135–156 (1906)
7. Bartlett, M.S.: Contingency table interactions. Supplement to the Journal of the Royal Statistical Society **2**(2), 248–252 (1935)
8. Seneta, E.: Markov and the birth of chain dependence theory. International Statistical Review / Revue Internationale de Statistique **64**(3), 255–263 (1996)
9. Yang, L.: Visualizing frequent itemsets, association rules, and sequential patterns in parallel coordinates. In: Kumar, V., Gavrilova, M., Tan, C., LEcuyer, P. (eds.) Computational Science and Its Applications ICCSA 2003. Lecture Notes in Computer Science, vol. 2667, pp. 21–30. Springer, Berlin, Germany (2003). http://dx.doi.org/10.1007/3-540-44839-X_3
10. Lafferty, J.D., McCallum, A., Pereira, F.C.N.: Conditional random fields: Probabilistic models for segmenting and labeling sequence data. In: Proceedings of the Eighteenth International Conference on Machine Learning. ICML '01, pp. 282–289. Morgan Kaufmann Publishers Inc., San Francisco, CA, USA (2001). http://dl.acm.org/citation.cfm?id=645530.655813
11. Inselberg, A.: The plane with parallel coordinates. The Visual Computer **1**, 69–91 (1985)
12. Rosario, G.E., Rundensteiner, E.A., Brown, D.C., Ward, M.O., Huang, S.: Mapping nominal values to numbers for effective visualization. Information Visualization **3**(2), 80–95 (2004)
13. Bendix, F., Kosara, R., Hauser, H.: Parallel sets: Visual analysis of categorical data. In: Information Visualization, 2005. INFOVIS 2005. IEEE Symposium On, pp. 133–140 (2005). IEEE
14. Lind, M., Johansson, J., Cooper, M.: Many-to-many relational parallel coordinates displays. In: Proceedings of the 2009 13th International Conference Information Visualisation. IV '09, pp. 25–31. IEEE Computer Society, Washington, DC, USA (2009)
15. Claessen, J.H.T., van Wijk, J.J.: Flexible linked axes for multivariate data visualization. IEEE Transactions on Visualization and Computer Graphics **17**(12), 2310–2316 (2011)
16. Lu, L.F., Huang, M.L., Huang, T.-H.: A new axes re-ordering method in parallel coordinates visualization. In: Machine Learning and Applications (ICMLA), 2012 11th International Conference On, vol. 2, pp. 252–257 (2012)
17. Makwana, H., Tanwani, S., Jain, S.: Article: Axes re-ordering in parallel coordinate for pattern optimization. International Journal of Computer Applications **40**(13), 43–48 (2012). Published by Foundation of Computer Science, New York, USA
18. Fanea, E., Carpendale, S., Isenberg, T.: An interactive 3d integration of parallel coordinates and star glyphs. In: Information Visualization, 2005. INFOVIS 2005. IEEE Symposium On, pp. 149–156 (2005)
19. Johansson, J., Ljung, P., Jern, M., Cooper, M.: Revealing structure in visualizations of dense 2d and 3d parallel coordinates. Information Visualization (2006)
20. Kerren, A., Jusufi, I.: 3d kiviat diagrams for the interactive analysis of software metric trends. In: Proceedings of the 5th International Symposium on Software Visualization. SOFTVIS '10, pp. 203–204. ACM, New York, NY, USA (2010). http://doi.acm.org/10.1145/1879211.1879241
21. Schmidt, M., Alahari, K.: Generalized fast approximate energy minimization via graph cuts: alpha-expansion beta-shrink moves. In: Proceedings of the 2011 IEEE Conference on Uncertainty in Artificial Intelligence. UAI'11, pp. 653–660 (2011)
22. Berry, M., Phillips Jr., G.N.: Crystal structures of bacillus stearothermophilus adenylate kinase with bound Ap5A,Mg2+Ap5a, and Mn2+ Ap5A reveal an intermediate lid position and six coordinate octahedral geometry for bound Mg2+ and Mn2+. Prot. Str. Func. Gen. **32**, 276–288 (1998)
23. Gavel, O.Y., Bursakov, S.A., DiRocco, G., Trincao, J., Pickering, I.J., George, G.N., Calvete, J.J., Shnyrov, V.L., Brondino, C.D., Pereira, A.S., Lampreia, J., Tavares, P., Maura, J.J., Maura, I.: A new type of metal-binding site in cobalt- and zinc-containing adenylate kinases isolated from sulfate-reducers desulfovibrio gigas and desulfovibrio desulfuricans atcc 27774. Jour. Inorganic Bioc. **102**, 1380–1395 (2008)
24. Berry, M.B., Bae, E., Bilderback, T.R., Glaser, M., Philips Jr., G.N.: Crystal structure of ADP/AMP construct of eschericia coli adenylate kinase. PROTEINS **62**, 555–556 (2005)
25. Ray, W.C.: MAVL/StickWRLD: Visualizing protein sequence families to detect non-consensus features. Nucleic Acids Research **33** – **Web Server Issue**, 315–319 (2005)
26. Perrier, V., Burlacu-Miron, S., Bourgeois, S., Surewicz, W.K., Gilles, A.-M.: Genetically engineered zinc-chelating adenylate kinase from *Eschericia coli* with enhanced thermal stability. Journal of Biological Chemistry **273**, 19097–19101 (1998)
27. Sim, N.-L., Kumar, P., Hu, J., Henikoff, S., Schneider, G., Ng, P.C.: Sift web server: predicting effects of amino acid substitutions on proteins. Nucleic Acids Research **40**(Web-Server-Issue), 452–457 (2012)
28. Adzhubei, I., Jordan, D.M., Sunyaev, S.R.: Predicting functional effect of human missense mutations using PolyPhen-2. Current protocols in human genetics / editorial board, Jonathan L. Haines ... [et al.] **Chapter 7** (2013)
29. Ray, W.C.: MAVL/StickWRLD: Visually exploring relationships in nucleic-acid sequence alignments. Nucleic Acids Research **32** – **Web Server Issue**, 59–63 (2004)
30. Fisher, R.A.: On the interpretation of x2 from contingency tables, and the calculation of p. Journal of the Royal Statistical Society **85**(1), 87–94 (1922)
31. Ozer, H.G.: Residue associations in protein family alignments. PhD thesis, The Ohio State University (June 2008)

# List of Figures

Figure 1   **Typical biological "sequence" data containing both positional and dependency information.** Sequences from Archaeal tRNA genes (A) and several canonical models and representations of this family of sequences. (B) Consensus, which simply represents the family in terms of the most popular symbol found in each column. (C) shows a Position Specific Scoring Matrix (PSSM), in this case truncated to single digit precision, which encodes the marginal distribution of each symbol in each column (D) shows a Sequence Logo, which convolves the marginal weights from a PSSM, with an information-theoretic measure of the information available in each column, under an assumption of column-column independence.(E) shows a sensory representation of the PSSM which provides some benefits for visually evaluating whether a candidate sequence fits the residue distribution of the training data. None of these representations provide any information regarding dependencies between either their columns, or between specific residues in specific columns. However, (E) provides a graphical starting point for an improved representation that can convey this information.

Figure 2   **The multigraph/metagraph structure underlying a GPM.** (A) Each position in the sequence, or distinct feature in the set, can be modeled as a node, while each observed category present at a location or feature, can be modeled as a subnode of that node. The weight of each subnode encodes the probability of finding that subnode's category in the training data, in that position. (B) Between every pair of nodes, there exists a complete bipartite graph of (potential) edges from the subnodes of one node to the other. Each edge encodes the probability of that connected pair of subnodes occurring in the training data. While it is easy to build this structure from the training data, it is almost always computationally intractable to use it to build a functional GPM. To create a tractably trainable GPM, the possible edges in (B) (and all other possible edges between each pair of columns) must be reduced to only the edges representing functionally important dependencies in the data.

Figure 3   **Extracting a simplified dependency structure to build a tractably trainable GPM.** To overcome the intractability shown in Figure 2:B, we need to simplify the edge structure of the resulting complete multi/metagraph such that it contains only the "most important" edges representing dependencies in the training data. Here we have shown a subset of the most important dependencies present in the data shown in Figure 1. While edge weights are not shown here, it is important in a working interface to provide the user with edge-weight information, and to avoid arbitrarily filtering edges based on their magnitude. To a biological end-user, small edges between infrequently occurring subnodes can be as important as larger edges between common subnodes, depending on the features they connect. Edges are colored based on disjoint subnetworks of dependencies.

Figure 4   **The bulge-helix-bulge structure targeted by an archaeal tRNA intron endonuclease is responsible for the data shown in Figure 1.** The exon positions are indicated by filled blocks while the intron positions are indicated by open blocks. The sequence consensus for each position is indicated beside its corresponding block. Position numbers correspond to the data shown in Figure 1.

Figure 5 **The simplified dependency structure found in the data from Figure 1, and shown in Figure 3, cast into the biological context of the molecular family from which the data was derived.** The aligned PSSM and interpositional dependencies for a sequence family identified by MAVL/StickWRLD correspond to a GPM where each possible base in each position represents to a node, and dependencies form edges. In this representation of a portion of the endonuclease target, the color of each node represents the base identity, and the size represents the frequency distribution of that base at that location in the sequence. Important positive dependencies are shown as black edges, and important negative dependencies are shown as dashed light-red edges. The "X shaped" dependencies in the lower stem correspond to the Watson-Crick interactions of a stem-loop structure. The dependencies within the bulge, and between the central stem and the bulge, are non-Watson-Crick, and are completely lost by other modeling methods. A few additional edges that are implied by the data are shown here, that could not be shown without overly cluttering Figure 3. **Fundamentally, the universal domain need is for a method of producing figures with similar information, without the need for significant manual intervention.**

Figure 6 **Wrapping categorical parallel-coordinate axes around a cylinder.** Arranging parallel coordinates axes around a cylinder, enables the complete graph of each feature vector to be displayed on the axes, rather than just one particular spanning walk. Using fixed vertical positions for each category and using scaled glyphs to represent node weights, enables the simultaneous display of all of the marginal probabilities of each category in each position, and all of the joint probabilities of every pairwise combination. We do not claim that this figure is visually tractable in this form, only that it does contain the features required for building useful GPMs.

Figure 7 **By exploring, filtering, and manually eliminating or saving various dependencies by brushing, a significantly simplified picture of the dependency network emerges.** Successive refinement from the raw dependency data shown in Figure 6, to a computationally tractable dependency structure for a CRF that enables accurate identification of other members of the sequence family. A) By applying the reduction of the displayed data to only the unexpected residuals, Figure 6 becomes much more sparse. B) Applying threshold filters to the magnitude of the residuals, further reduces the visual complexity of the model and simultaneously decreases the likelihood of overfitting the data with the CRF model, and brings the dependency network closer to being computationally tractable. C) Finally applying statistical filters, and manual editing of the dependency structure, results in a CRF dependency model that captures the important family sequence signatures. It is also relatively easy to browse and understand in the interactive interface, despite casting the parallel axes in a volume rather than a plane.

Figure 8 **Birds-eye views of the dependency network in ADK while being explored by a user to identify a subset suitable for building a CRF.** Views from StickWRLD being used to refine a GPM to identify the most critical determinants of catalytic activity in the Adenylate Kinase lid domain. A) shows an overhead view of the 300-column, 21-category dependency network after it has already been filtered down to relatively large residuals. B), C) and D) show successive refinements using a statistical threshold cutoff. By the time $p = 0.001$ has been reached in D), the majority of the interesting interactions have been lost.

Figure 9 **An adequately refined view of the ADK dependencies for building a CRF.** Eliminating the majority of columns with no dependencies from Figure 8:C, focuses attention on the known interacting tetrad of residues in the lid domain, and on other residues that show dependencies with these. Our 4, 6 and 12 dependency CRF models were derived from this view.

Figure 10   **Visualized CRF of a known polyadenylation signal motif.** A StickWRLD view of the genomic sequence motif that governs "signal" based polyadenylation. Colors and categories are as previously shown. The motif representation starts in the back of the cylinder and proceeds counterclockwise. It is relatively easily modeled as 4 "don't care" positions with no significant base preference, followed by 2 A bases (red balls), a single T base (blue ball), and then three more A bases. There are a few very small residuals that attain statistical significance, but the marginal distributions dominate the motif, allowing it to be found using both PSSM and HMM methods.

Figure 11   **Visualized implied CRF from a misaligned polyadenylation signal model.** Visualizing the "non signal" sequence regions, we see a strong disposition towards A and T bases in the marginal distributions, but no overwhelming pattern. PSSM and HMM methods fail to identify a pattern in these sequences, however, the fact that we do not restrict StickWRLD to only sequentially adjacent dependencies lets us see that there is a curious "echoing" pattern of dependencies between $T_15$ and $A_17$, $T_16$ and $A_18$, $T_17$ and $A_19$, and $T_18$ and $A_20$. This echo extends further, at lower residual and significance thresholds. This echo is a highly suggestive fingerprint of a misalignment in the data. We are seeing an interdependency between a T and an A base, 2 bases apart, occurring in several subsets of the input data, each shifted slightly from the next. Using StickWRLD to interact with and realign the data, we arrive at Figure 12.

Figure 12   **Corrected CRF for the apparently signal-less polyadenylation signals.** After realignment, we can see that the "non signal" polyadenylation signals actually do have a strongly conserved pattern of residues, but that unlike the "signal" motif, the motif also possesses significant dependencies. Notably, rather than a pair of A residues followed by a single $T$ (as seen in the "signal" motif), these sequences possess a single $A$ residue, followed with almost equal probability by an $A$ or a $T$ residue. That $A$ or $T$ residue strongly influences the identity of the subsequent residue – if the first is a $T$, then the second is also a $T$, if the first is an $A$, then the second is also an $A$. This variable pair of residues is then followed by, as in the "signal" motif, a trio of $A$ residues ending the motif. Several other dependencies also show up. The interdependencies visualized here, are why PSSM and HMM models have failed to identify an alignment in, or adequately model this "non signal" signaling motif.

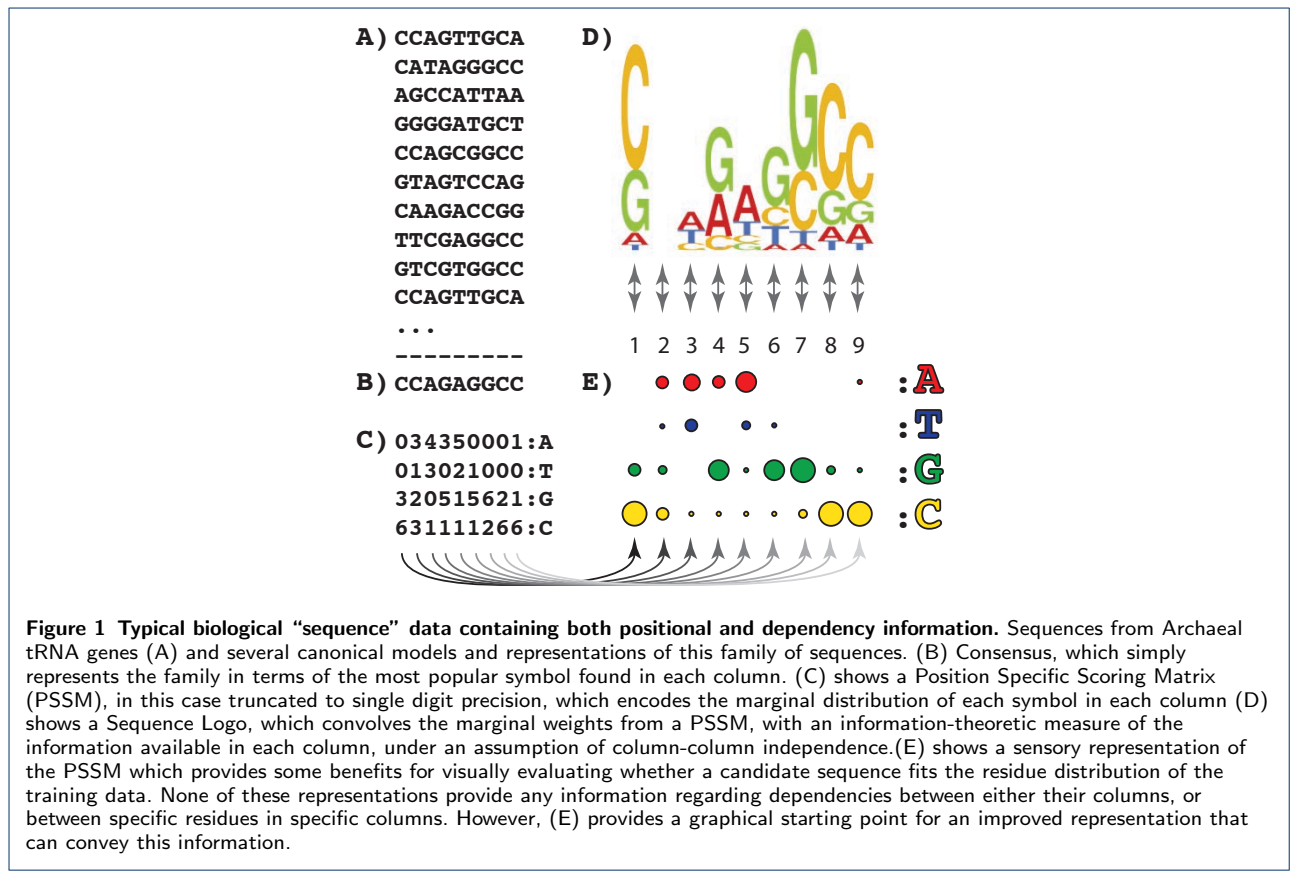

**Figure 1 Typical biological "sequence" data containing both positional and dependency information.** Sequences from Archaeal tRNA genes (A) and several canonical models and representations of this family of sequences. (B) Consensus, which simply represents the family in terms of the most popular symbol found in each column. (C) shows a Position Specific Scoring Matrix (PSSM), in this case truncated to single digit precision, which encodes the marginal distribution of each symbol in each column (D) shows a Sequence Logo, which convolves the marginal weights from a PSSM, with an information-theoretic measure of the information available in each column, under an assumption of column-column independence.(E) shows a sensory representation of the PSSM which provides some benefits for visually evaluating whether a candidate sequence fits the residue distribution of the training data. None of these representations provide any information regarding dependencies between either their columns, or between specific residues in specific columns. However, (E) provides a graphical starting point for an improved representation that can convey this information.

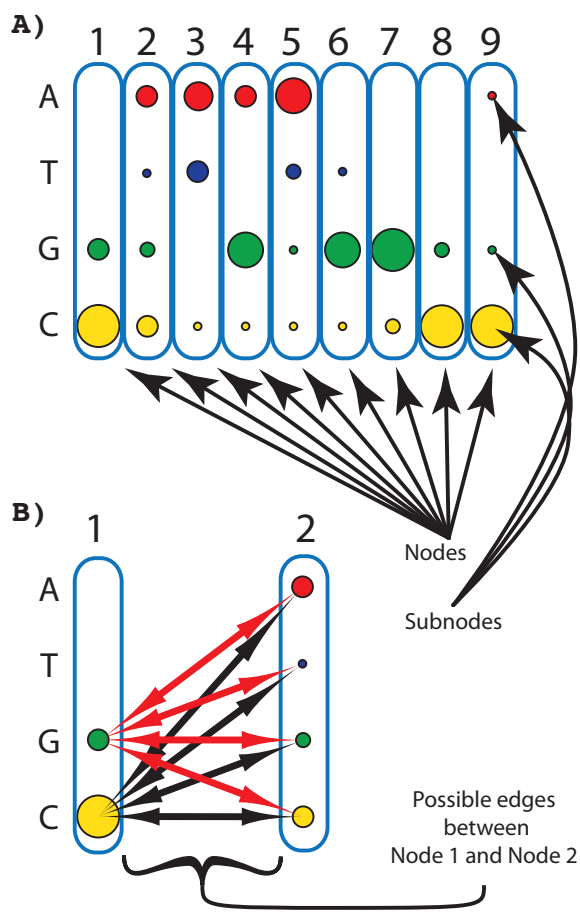

**Figure 2 The multigraph/metagraph structure underlying a GPM.** (A) Each position in the sequence, or distinct feature in the set, can be modeled as a node, while each observed category present at a location or feature, can be modeled as a subnode of that node. The weight of each subnode encodes the probability of finding that subnode's category in the training data, in that position. (B) Between every pair of nodes, there exists a complete bipartite graph of (potential) edges from the subnodes of one node to the other. Each edge encodes the probability of that connected pair of subnodes occurring in the training data. While it is easy to build this structure from the training data, it is almost always computationally intractable to use it to build a functional GPM. To create a tractably trainable GPM, the possible edges in (B) (and all other possible edges between each pair of columns) must be reduced to only the edges representing functionally important dependencies in the data.

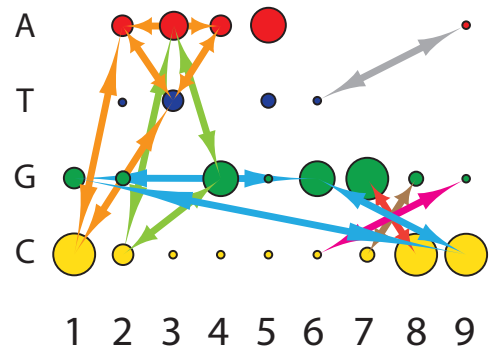

**Figure 3 Extracting a simplified dependency structure to build a tractably trainable GPM.** To overcome the intractability shown in Figure 2:B, we need to simplify the edge structure of the resulting complete multi/metagraph such that it contains only the "most important" edges representing dependencies in the training data. Here we have shown a subset of the most important dependencies present in the data shown in Figure 1. While edge weights are not shown here, it is important in a working interface to provide the user with edge-weight information, and to avoid arbitrarily filtering edges based on their magnitude. To a biological end-user, small edges between infrequently occurring subnodes can be as important as larger edges between common subnodes, depending on the features they connect. Edges are colored based on disjoint subnetworks of dependencies.

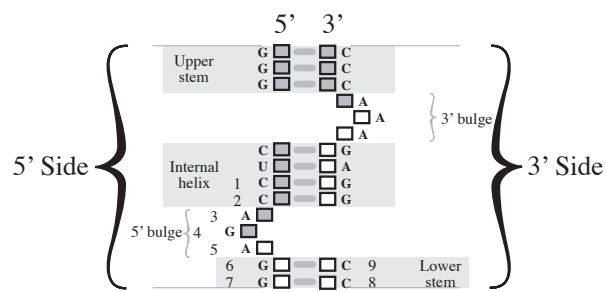

**Figure 4 The bulge-helix-bulge structure targeted by an archaeal tRNA intron endonuclease is responsible for the data shown in Figure 1.** The exon positions are indicated by filled blocks while the intron positions are indicated by open blocks. The sequence consensus for each position is indicated beside its corresponding block. Position numbers correspond to the data shown in Figure 1.

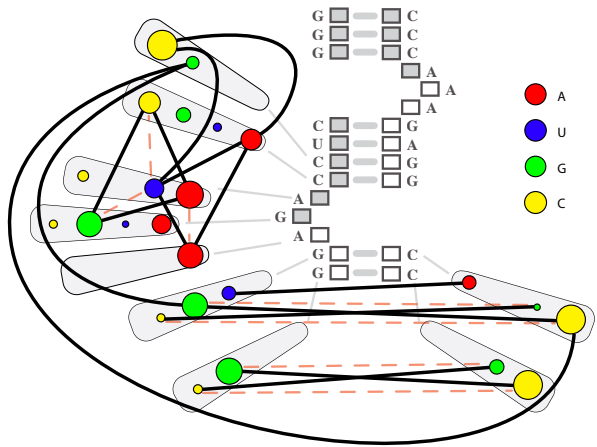

**Figure 5 The simplified dependency structure found in the data from Figure 1, and shown in Figure 3, cast into the biological context of the molecular family from which the data was derived.** The aligned PSSM and interpositional dependencies for a sequence family identified by MAVL/StickWRLD correspond to a GPM where each possible base in each position represents to a node, and dependencies form edges. In this representation of a portion of the endonuclease target, the color of each node represents the base identity, and the size represents the frequency distribution of that base at that location in the sequence. Important positive dependencies are shown as black edges, and important negative dependencies are shown as dashed light-red edges. The "X shaped" dependencies in the lower stem correspond to the Watson-Crick interactions of a stem-loop structure. The dependencies within the bulge, and between the central stem and the bulge, are non-Watson-Crick, and are completely lost by other modeling methods. A few additional edges that are implied by the data are shown here, that could not be shown without overly cluttering Figure 3. **Fundamentally, the universal domain need is for a method of producing figures with similar information, without the need for significant manual intervention.**

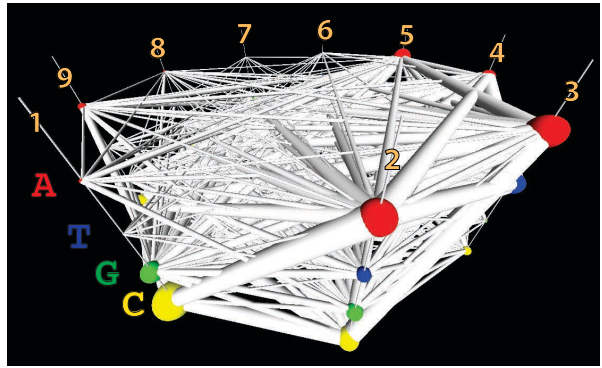

**Figure 6 Wrapping categorical parallel-coordinate axes around a cylinder.** Arranging parallel coordinates axes around a cylinder, enables the complete graph of each feature vector to be displayed on the axes, rather than just one particular spanning walk. Using fixed vertical positions for each category and using scaled glyphs to represent node weights, enables the simultaneous display of all of the marginal probabilities of each category in each position, and all of the joint probabilities of every pairwise combination. We do not claim that this figure is visually tractable in this form, only that it does contain the features required for building useful GPMs.

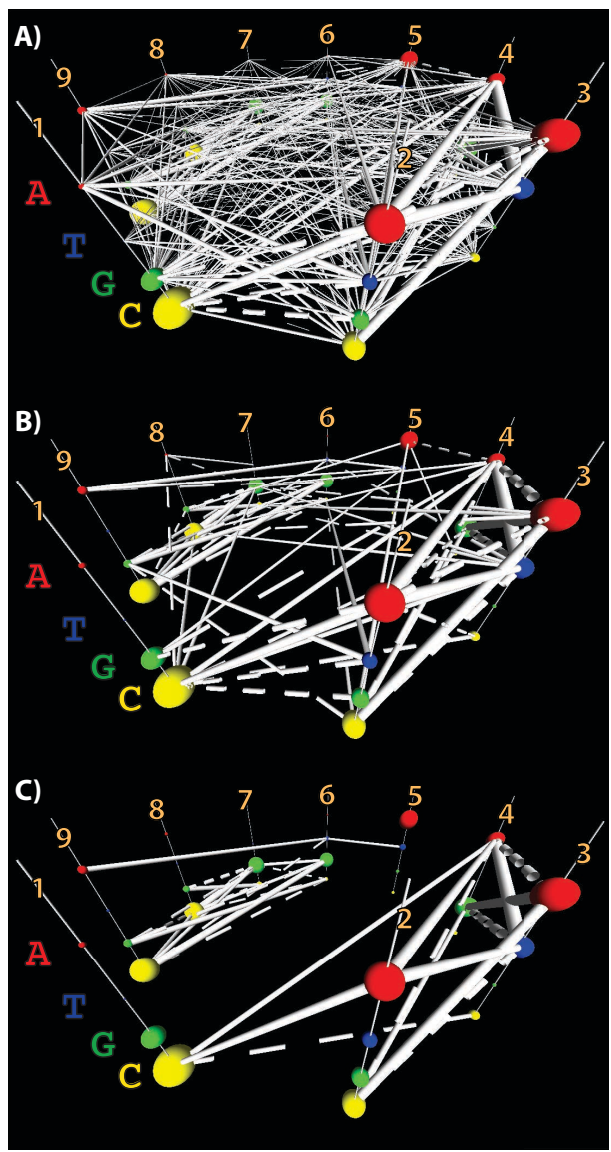

**Figure 7 By exploring, filtering, and manually eliminating or saving various dependencies by brushing, a significantly simplified picture of the dependency network emerges.** Successive refinement from the raw dependency data shown in Figure 6, to a computationally tractable dependency structure for a CRF that enables accurate identification of other members of the sequence family. A) By applying the reduction of the displayed data to only the unexpected residuals, Figure 6 becomes much more sparse. B) Applying threshold filters to the magnitude of the residuals, further reduces the visual complexity of the model and simultaneously decreases the likelihood of overfitting the data with the CRF model, and brings the dependency network closer to being computationally tractable. C) Finally applying statistical filters, and manual editing of the dependency structure, results in a CRF dependency model that captures the important family sequence signatures. It is also relatively easy to browse and understand in the interactive interface, despite casting the parallel axes in a volume rather than a plane.

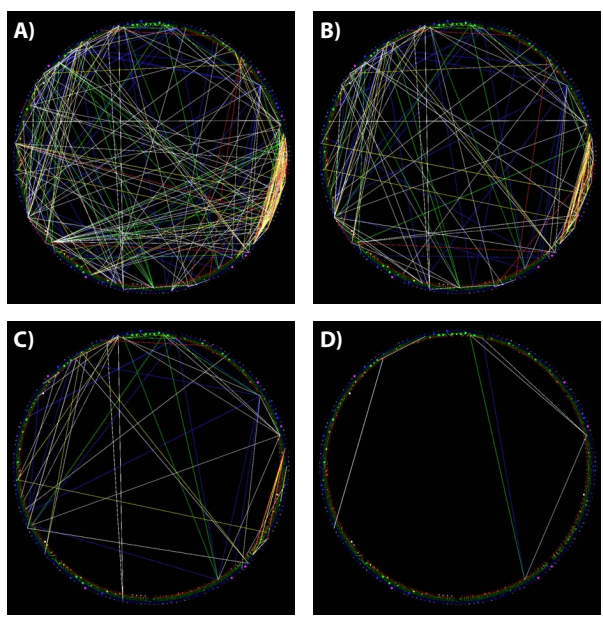

**Figure 8 Birds-eye views of the dependency network in ADK while being explored by a user to identify a subset suitable for building a CRF.** Views from StickWRLD being used to refine a GPM to identify the most critical determinants of catalytic activity in the Adenylate Kinase lid domain. A) shows an overhead view of the 300-column, 21-category dependency network after it has already been filtered down to relatively large residuals. B), C) and D) show successive refinements using a statistical threshold cutoff. By the time $p = 0.001$ has been reached in D), the majority of the interesting interactions have been lost.

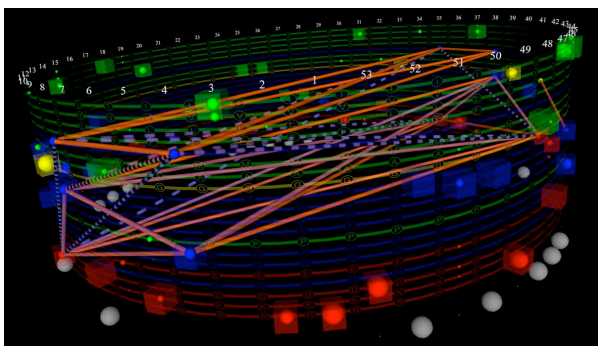

**Figure 9 An adequately refined view of the ADK dependencies for building a CRF.** Eliminating the majority of columns with no dependencies from Figure 8:C, focuses attention on the known interacting tetrad of residues in the lid domain, and on other residues that show dependencies with these. Our 4, 6 and 12 dependency CRF models were derived from this view.

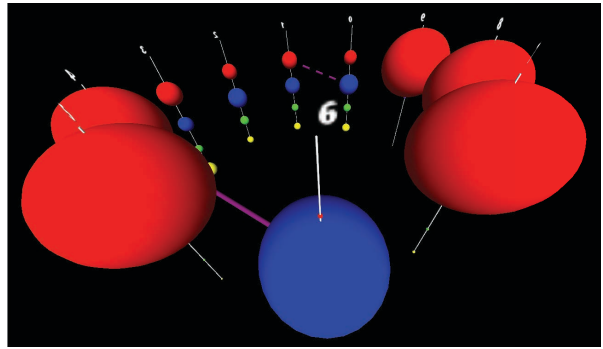

**Figure 10 Visualized CRF of a known polyadenylation signal motif.** A StickWRLD view of the genomic sequence motif that governs "signal" based polyadenylation. Colors and categories are as previously shown. The motif representation starts in the back of the cylinder and proceeds counterclockwise. It is relatively easily modeled as 4 "don't care" positions with no significant base preference, followed by 2 A bases (red balls), a single T base (blue ball), and then three more A bases. There are a few very small residuals that attain statistical significance, but the marginal distributions dominate the motif, allowing it to be found using both PSSM and HMM methods.

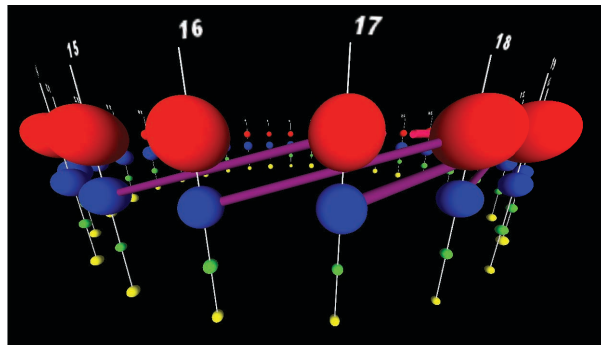

**Figure 11 Visualized implied CRF from a misaligned polyadenylation signal model.** Visualizing the "non signal" sequence regions, we see a strong disposition towards A and T bases in the marginal distributions, but no overwhelming pattern. PSSM and HMM methods fail to identify a pattern in these sequences, however, the fact that we do not restrict StickWRLD to only sequentially adjacent dependencies lets us see that there is a curious "echoing" pattern of dependencies between $T_15$ and $A_17$, $T_16$ and $A_18$, $T_17$ and $A_19$, and $T_18$ and $A_20$. This echo extends further, at lower residual and significance thresholds. This echo is a highly suggestive fingerprint of a misalignment in the data. We are seeing an interdependency between a T and an A base, 2 bases apart, occurring in several subsets of the input data, each shifted slightly from the next. Using StickWRLD to interact with and realign the data, we arrive at Figure 12.

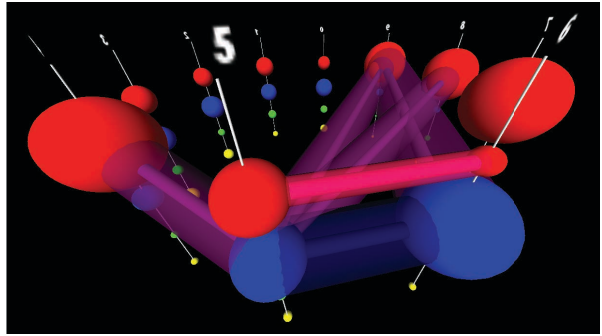

**Figure 12 Corrected CRF for the apparently signal-less polyadenylation signals.** After re-alignment, we can see that the "non signal" polyadenylation signals actually do have a strongly conserved pattern of residues, but that unlike the "signal" motif, the motif also possesses significant dependencies. Notably, rather than a pair of A residues followed by a single $T$ (as seen in the "signal" motif), these sequences possess a single $A$ residue, followed with almost equal probability by an $A$ or a $T$ residue. That $A$ or $T$ residue strongly influences the identity of the subsequent residue – if the first is a $T$, then the second is also a $T$, if the first is an $A$, then the second is also an $A$. This variable pair of residues is then followed by, as in the "signal" motif, a trio of $A$ residues ending the motif. Several other dependencies also show up. The interdependencies visualized here, are why PSSM and HMM models have failed to identify an alignment in, or adequately model this "non signal" signaling motif.

# List of Tables

**Table 1** *B. subtilis* mutants and activity fold changes. Residues changed from wild-type are indicated in bold.

The relative activities of these mutants show that not only the identities of the residues in each position, but also the relationships between these residues play a key role in enzyme activity. Position 24, for example, has an almost equal probability of containing a **C**, or **D** residue, across the ADK family. The functional consequences of changing a **C** to a **D** in a specific protein however, must be calculated in the context of the other residues in that specific protein for which relational dependencies exist. Our assays show that activity correlates well with the predictions from a CRF defined using the network formed from these relationships. The **Di** mutant retains activity, only slightly impaired from wild-type. The **Tetra** mutant shows barely detectable activity. The **Hexa** mutant recovers a significant amount of activity, but remains an order of magnitude less active than wild-type. CD thermal denaturation shows little difference in stability between the wild-type and **Di** mutants and only a small destabilization in the **Tetra** and **Hexa** mutants.

All of these activity changes agree with predictions for the modified *B. subtilis* sequence, by a CRF defined by the interdependencies between the residues of these motifs – with one caveat. If only these residues are used to define the CRF, it predicts that the **Hexa** mutant will have better activity than the wild-type protein.

This caveat highlights the danger of assuming that only the very strongest co-evolutions are necessary to define an adequate CRF. The CRF defined with the 6 residues most obviously involved (CRF6), fails to evaluate those residues *in the context of the rest of the specific B. subtilis residues in the protein*. Because the hydrophilic residue motif is more prevalent in the training set, the CRF predicts that a mutant containing it, will be more likely to be functional. This failure is exactly why network models of interdependency are critical for developing accurate predictive methods for protein sequence → function.

|        |   | position |    |    |    |    |         |           |            |            |
|--------|---|---|----|----|----|----|---------|-----------|------------|------------|
| mutant | 4 | 7 | 10 | 24 | 27 | 29 | fold | CRF4 | CRF6 | CRF12 |
| BsADK  | C | C | T  | C  | D  | G  | 1       | 1         | 1          | 1          |
| Chim   | C | C | T  | **D** | **T** | G  | $-\infty$ | $10^{-68}$ | $10^{-150}$ | $10^{-248}$ |
| Tetra  | **H** | **S** | T | **D** | **T** | G | *inactive*$++$ | $10^{1}$ | $10^{-70}$ | $10^{-146}$ |
| Di     | C | C | **R** | C | D | **E** | *normal*$-$ | $10^{1}$ | $10^{-2}$ | $10^{-2}$ |
| Hexa   | **H** | **S** | **R** | **D** | **T** | **E** | *normal*$--$ | $10^{3}$ | $10^{2}$ | $10^{-46}$ |